\documentclass[12pt,preprint]{aastex}

\def \egret {EGRET}

\def \phcmsec{\hbox{photons cm$^{-2}$ s$^{-1}$}}

\def \gray {$\gamma$-ray }

\def \source {\hbox{3C~454.3}}

\slugcomment{To appear in ApJ}

\shorttitle{AGILE detection of 3C~454.3}
\shortauthors{Vercellone et al.}

\begin{document}
\title{AGILE detection of a strong gamma-ray flare from the blazar 3C~454.3}
\author{S.~Vercellone\altaffilmark{1,*}, A.W.~Chen\altaffilmark{1,2},
  A.~Giuliani\altaffilmark{1}, A.~Bulgarelli\altaffilmark{3},
  I.~Donnarumma\altaffilmark{4}, I.~Lapshov\altaffilmark{4}, 
  M.~Tavani\altaffilmark{4,5},
  A.~Argan\altaffilmark{4}, G.~Barbiellini\altaffilmark{6},
  P.~Caraveo\altaffilmark{1}, V.~Cocco\altaffilmark{4},
  E.~Costa\altaffilmark{4},
  F.~D'Ammando\altaffilmark{4,5},
  E.~Del Monte\altaffilmark{4}, G.~De Paris\altaffilmark{4},
  G.~Di Cocco\altaffilmark{3},
  Y.~Evangelista\altaffilmark{4},
  M.~Feroci\altaffilmark{4}, M.~Fiorini\altaffilmark{1},
  T.~Froysland\altaffilmark{2,5},
  F.~Fuschino\altaffilmark{3}, M.~Galli\altaffilmark{7},
  F.~Gianotti\altaffilmark{3}, C.~Labanti\altaffilmark{3}, 
  F.~Lazzarotto\altaffilmark{4}, 
  P.~Lipari\altaffilmark{8}, F.~Longo\altaffilmark{6}, 
  M.~Marisaldi\altaffilmark{3},
  F.~Mauri\altaffilmark{9}, S.~Mereghetti\altaffilmark{1},
  A.~Morselli\altaffilmark{10}, L.~Pacciani\altaffilmark{4},
  A.~Pellizzoni\altaffilmark{1}, F.~Perotti\altaffilmark{1},
  P.~Picozza\altaffilmark{10}, M.~Prest\altaffilmark{11},
  G.~Pucella\altaffilmark{4}, M.~Rapisarda\altaffilmark{12},
  P.~Soffitta\altaffilmark{4}, M.~Trifoglio\altaffilmark{3},
  A.~Trois\altaffilmark{4}, E.~Vallazza\altaffilmark{6},
  V.~Vittorini\altaffilmark{5}, A.~Zambra\altaffilmark{1},
  D.~Zanello\altaffilmark{8},
  C.~Pittori\altaffilmark{13}, F.~Verrecchia\altaffilmark{13},
  D.~Gasparrini\altaffilmark{13}, S.~Cutini\altaffilmark{13},
  P.~Giommi\altaffilmark{13}, L.A.~Antonelli\altaffilmark{13},
  S.~Colafrancesco\altaffilmark{13}, L.~Salotti\altaffilmark{14}
}
\altaffiltext{1}{INAF/IASF--Milano, Via E.~Bassini 15, I-20133 Milano, Italy}
\altaffiltext{2}{CIFS--Torino, Viale Settimio Severo 3, I-10133, Torino, Italy}
\altaffiltext{3}{INAF/IASF--Bologna, Via Gobetti 101, I-40129 Bologna, Italy}
\altaffiltext{4}{INAF/IASF--Roma, Via del Fosso del Cavaliere 100, 
  I-00133 Roma, Italy}
\altaffiltext{5}{Dip. di Fisica, Univ. ``Tor Vergata'', Via della Ricerca 
  Scientifica 1, I-00133 Roma, Italy}
\altaffiltext{6}{Dip. di Fisica and INFN, Via Valerio 2, I-34127 Trieste, Italy}
\altaffiltext{7}{ENEA--Bologna, Via Biancafarina 2521, I-40059 Medicica (BO), 
  Italy}
\altaffiltext{8}{INFN--Roma ``La Sapienza'', Piazzale A. Moro 2, I-00185 Roma, 
  Italy}
\altaffiltext{9}{INFN--Pavia, Via Bassi 6, I-27100 Pavia, Italy}
\altaffiltext{10}{INFN--Roma ``Tor Vergata'', Via della Ricerca Scientifica 1, 
  I-00133 Roma, Italy}
\altaffiltext{11}{Dip. di Fisica, Univ. dell'Insubria, Via Valleggio 11, 
  I-22100 Como, Italy}
\altaffiltext{12}{ENEA--Roma, Via E. Fermi 45, I-00044 Frascati (Roma), Italy}
\altaffiltext{13}{ASI--ASDC, Via G. Galilei, I-00044 Frascati (Roma), Italy}
\altaffiltext{14}{ASI, Viale Liegi 26 , I-00198 Roma, Italy}
\altaffiltext{*}{Email: \texttt{stefano@iasf-milano.inaf.it}}

\begin{abstract}
We report the first blazar detection by the AGILE satellite.
AGILE detected \source{} during a 
period of strongly enhanced optical emission in July 2007.
AGILE observed the source with a dedicated repointing 
during the period 2007 July 24--30 with its two co-aligned imagers,
the Gamma-Ray Imaging Detector and the hard X-ray imager 
Super-AGILE sensitive in the
30~MeV--50~GeV and 18--60 keV, respectively.
Over the entire period, AGILE detected \gray emission 
from \source{} at a significance level of
13.8-$\sigma$ with an average flux (E$>$100~MeV) of 
$(280 \pm 40) \times 10^{-8}$\,\phcmsec. 
The \gray flux appears to be variable towards the end of the
observation.
No emission was detected by Super-AGILE in the energy range 20--60~keV, 
with a 3-$\sigma$ upper limit of $2.3 \times 10^{-3}$\,\phcmsec.
The \gray flux level of \source{} detected by AGILE is 
the highest ever detected for this quasar and among the most intense 
\gray fluxes ever detected from Flat Spectrum Radio Quasars.
\end{abstract}

        \section{Introduction \label{3c454:introduction}}

Among active galactic nuclei (AGNs), blazars
show strong flux variability at almost all frequencies of their
spectral energy distributions (SED). The Energetic Gamma Ray Experiment 
Telescope (\egret) instrument  on board the 
{\it Compton Gamma-Ray Observatory} ({\it CGRO})
detected, above 30~MeV, several  AGNs, establishing the blazars as 
a class of \gray sources \citep{Hartman1999:3eg}.
Gamma-ray blazars are characterized by high variability on different 
timescales. At energies above 30 MeV, variability has been detected 
on timescales
from one day (e.g. PKS~1622$-$297, \citealt{Zhang2002:pks1622}) to one month
(e.g. PKS~0208$-$512, \citealt{Montigny1995:pks0208}).

The flat--spectrum radio quasar (FSRQ) \source{}
(PKS~2251$+$158; $z=0.859$) was detected by EGRET in 1992
during an intense \gray flaring episode 
\citep{Hartman1992:3C454iauc, Hartman1993:3C454_EGRET}
when the blazar flux ($E > 100$\,MeV) was observed to vary within the range
$(0.4-1.4) \times 10^{-6}$\,photons\,cm$^{-2}$\,s$^{-1}$. In 1995, a 2-week 
EGRET campaign detected a \gray flux $< 1/5$ of its historical maximum
\citep{Aller1997:3C454_EGRET}.
\citet{Vercellone2004:duty}, analyzing the 67 EGRET \gray blazars
by defining an activity index $\psi$,
singled out \source{} as the source with the highest activity index
($\psi=0.033 \times 10^{-7}$\,cm$^{-2}$\,s$^{-1}$) among EGRET blazars.

In 2005, \source{} displayed major flaring activity in almost 
all energy bands (see \citealt{Giommi2006:3C454_Swift}).
In the optical, it reached $R=12.0$\,mag 
\citep{Villata2006:3C454_WEBT0405} 
and it was detected by INTEGRAL at a flux\footnote{Assuming
a Crab-like spectrum.} level of
$\sim 3 \times 10^{-2}$\,photons\,cm$^{-2}$\,s$^{-1}$ in the 3--200~keV 
energy band \citep{Pian2006:3C454_Integral}. 
The radio flux began increasing
right after the optical flare, and reached a maximum about 
a year later \citep{Villata2007:3C454_WEBT0506}.
Unfortunately, at that time no high \gray instrument 
was operational. As pointed out by \citet{Pian2006:3C454_Integral},
simultaneous observations of FSRQs by means of high-energy instruments
are crucial to precisely locate the inverse Compton peak and
to constrain the physical mechanisms operating in AGN flares.

In July 2007, \source{} showed
significant activity with strong optical flaring 
episodes (G. Tosti, 2007, private communication),
reaching $R=12.8$\,mag\footnote{
http://users.utu.fi/$\sim$kani/1m/index.html}.
This time, the recently launched AGILE satellite was
performing its Science Performance Verification Phase (SVP)
and devoted a week to the observation of \source{},
joining a multi-frequency campaign 
reacting to the enhanced optical activity

The AGILE satellite, 
a mission \citep{Tavani2006:AGILE_SPIE} of the Italian Space 
Agency (ASI) devoted to high-energy
astrophysics, is currently the only
space mission capable of observing cosmic sources simultaneously
in the energy bands 18--60~keV and 30~MeV -- 50~GeV. The satellite
was launched on 2007 April 23 by the Indian
PSLV-C8 rocket from the Satish Dhawan Space Center SHAR,
Sriharikota. The AGILE equatorial orbit (average height 540 km,
inclination angle 2.5 degrees) is ideal for
high-energy observations because of its 
low-background environment.
After the post-launch Commissioning Phase (2007 May-June), AGILE
started its SVP. 

In this Letter we present an analysis of the AGILE data
obtained during the \source{} observations.
Preliminary results were communicated in 
\citet{Vercellone2007:atel1160} and \citet{Bulgarelli2007:atel1167}.
The results of a multiwavelength campaign on \source{} and
the theoretical modelling will be presented in a forthcoming paper.
Throughout this paper the quoted uncertainties are given at the 
1--$\sigma$ level, unless otherwise stated.

        \section{The AGILE Instrument} \label{3c454:instrument}

The AGILE satellite (M.\ Tavani et al.\ 2008a, in preparation) has a 
3-axis attitude stabilization with fixed solar panels. Because of 
the solar panel constraints, AGILE must mantain the 
Instrument axis oriented at 90$^{\circ}$ from the Sun at all times. 
Thus, the locus of allowed pointings forms a great circle in the sky.
The AGILE scientific Instrument (M.\ Tavani et al.\ 2008b, in preparation) 
is very compact and combines four
active detectors yielding broad-band coverage from hard X-rays
to gamma-rays.
Gamma-ray detection is obtained by the combination of a Silicon Tracker, 
a Mini-Calorimeter and an  Anticoincidence System; these three
detectors form the AGILE Gamma-Ray Imaging Detector (GRID). 
The Silicon Tracker
\citep{Prest2003:agile_st,Barbiellini2001:agile_st} and the
on-board trigger logic (A.\ Argan et al. 2008, in preparation) 
are optimized for
\gray imaging in the 30~MeV -- 50~GeV energy band.
A non-imaging CsI detector
(Mini--Calorimeter, MCAL) is positioned under the Silicon
Tracker and is sensitive in the 0.4--100~MeV energy band
\citep{Labanti2006:agile_mcal}. 
A co-aligned coded-mask hard X-ray imager 
\citep[SA,][]{Feroci2007:agile_sa,Costa2001:agile_sa}
ensures simultaneous coverage in the 18--60~keV energy band.
A segmented Anti-Coincidence System (ACS) made of a plastic
shield surrounds all active detectors
\citep{Perotti2006:agile_ac}.
Table~\ref{3c454:tab:tab1}
provides the relevant parameters for both AGILE imaging detectors
(AGILE-GRID and Super-AGILE)
as computed on the basis of 
on ground calibrations, GRID beam tests, and
extensive Monte-Carlo simulations.

A crucial part of the GRID event processing is provided
by the on-board Data Handling unit (\citealt{Giuliani2006:kalmex};
A.\ Argan et al.\ 2008, in preparation). 
The on-ground data processing handles the GRID and Super-AGILE 
data through dedicated software developed by the instrument Teams 
and integrated into an automatic pipeline system at the 
ASI Science Data Center (ASDC\footnote{http://agile.asdc.asi.it/}). 
The reduction software runs both at the ASDC and at the 
Instrument Institutes.

The SVP is aimed at testing and optimizing the overall
scientific performance of the Instrument, including in-flight
calibrations. At the time of writing, the SVP has just 
been completed, and only a partial analysis of the
in-flight calibration data is available, therefore
we restrict our analysis to the E$>100$\,MeV energy range.
However, the successful
observations and tests carried out during the Commissioning Phase
allow us to reliably process the data and obtain
preliminary scientific results. 
%

        \section{AGILE observation of \source{} } \label{3c454:pointing}

At the epoch of the \source{} optical flare (2007 July 19--21) the
closest pointing position allowed by the solar panel
constraints ranged within 35$^{\circ}$--40$^{\circ}$ from the source.
Because of the remarkably large ($\sim 3$\,sr) field of view (FOV)
of the GRID and the successful
detection of the Vela pulsar at $\sim 55^{\circ}$ off-axis
obtained during the SVP, AGILE could
reliably study \source{} despite its large off-axis position.
The observation were performed between 2007 July 24 14:30 UT  and 2007 
July 30 11:40 UT, for a total pointing duration of $\sim 5.8$ days.

        \section{Data Reduction and Analysis} \label{3c454:dataanal}

Level--1 AGILE-GRID data were analyzed using the
AGILE Standard Analysis Pipeline. 
A first step aligns all data times to Terrestrial Time (TT)
and performs preliminary calculations. 
In a second step, an ad-hoc implementation of the Kalman Filter
technique (A.\ Giuliani et al., 2008 in preparation) is used
for track identification and event direction reconstruction in detector
coordinates.
Subsequently, a quality flag is assigned to each GRID event:
(G), (P), (S), and (L), depending on whether 
it is recognized as a \gray event,
a charged particle event, a single-track event, or
if its nature is uncertain, respectively.
Then, an AGILE log-file
is created, containing all the information relevant to the
computation of the exposure and live-time.
A third step creates the AGILE event files, 
excluding events flagged as particles.
This step also reconstructs the event direction in sky coordinates.

Once the above steps are completed, the 
AGILE Scientific Analysis Package can be run.
Counts, exposure, and Galactic background \gray maps are created with
a bin-size of $0.\!\!^{\circ}5 \times 0.\!\!^{\circ}5$\,,
for $E \ge 100$\,MeV.
To reduce the particle background contamination
we selected only events flagged as 
confirmed \gray events (\texttt{filtercode=5}), while all events 
collected during the South Atlantic Anomaly (SAA) were
rejected (\texttt{phasecode=18}).
We also rejected all the \gray events whose reconstructed
directions form angles with the satellite-Earth vector smaller
than 80$^{\circ}$ (\texttt{albrad=80}), 
reducing the \gray Earth Albedo contamination
by excluding regions within $\sim 10^{\circ}$ from the
Earth limb.
The most recent versions (\texttt{BUILD-12}) of the Calibration files,
which will be publicly available at the ASDC site,
and of the \gray diffuse emission model \citep{Giuliani2004:diff_model}
were used. 

Since a complete analysis of the
in-flight calibration is not available, yet, we adopted the following
procedure for the \gray flux determination.
We ran the AGILE Maximum Likelihood procedure (ALIKE) on the whole
observing period, in order to obtain a value for the average flux.
We ran ALIKE both on \source{} and on a long ($\sim 12$\,days) 
set of observations of Vela pulsar taken at 
approximately 23$^{\circ}$ off-axis.
Since Monte-Carlo simulations show that the effective area
and the point-spread function are similar for off-axis angles
in the 25$^{\circ}$--35$^{\circ}$ range,
we estimated the \source{} flux according to
\begin{equation}\label{3c454:equ:defflux}
  F_{\rm 3C~454.3}^{\rm \,est} = 
  \frac{F_{\rm 3C~454.3}^{\rm ALIKE}}{F_{\rm Vela}^{\rm ALIKE}}
  \times F_{\rm Vela}^{\rm 3EG}\,,
\end{equation}
where the terms with the ALIKE superscript are derived by ALIKE,
while those with the 3EG superscript come from the Third EGRET
Catalog \citep{Hartman1999:3eg}.

Super-AGILE observed \source{} for a total on-source net
exposure time of 250 ks. 
The source was observed at an off-axis position
varying from $0.\!^{\circ}7$ to $2.\!^{\circ}5$ along the SA
$X$--coordinate and from $-37.\!^{\circ}3$ to $-42.\!^{\circ}2$ along
the $Z$--coordinate, due to the Sun constraints on the AGILE pointing. 
This implies that the source could be observed
only by two of the four SA detectors, with imaging
capability in only one direction ($Z$) and an exposed area of about
1/8 of the on-axis value (see \citealt{Feroci2007:agile_sa} for a detailed
description of the geometry of the FoV of the
Super-AGILE detector).

        \section{Results} \label{3c454:results}

\begin{figure}[!h] 
    \includegraphics[angle=0,scale=0.40]{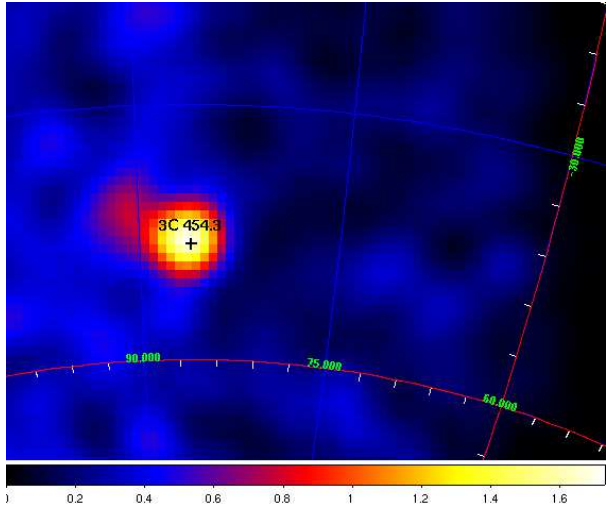}
    \caption[Map of 3C~454.3.]{Gaussian-smoothed counts map 
      ($\sim 38^{\circ} \times 28^{\circ}$)
      in Galactic coordinates integrated over the whole 
      observing period (2007 July 24 14:30 UT -- 2007 July 30 11:40 UT). 
      The cross symbol is located at the \source{}
      radio coordinates.
    \label{3c454:fig:map}}
\end{figure}

\begin{figure}[!h] 
    \includegraphics[angle=90,scale=0.55]{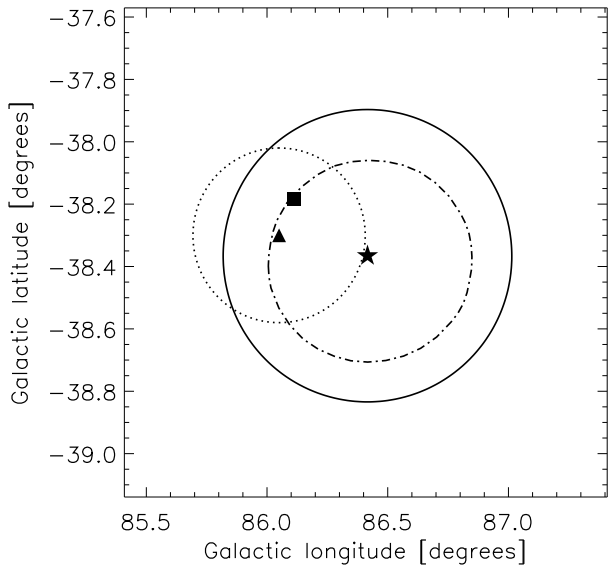}
    \caption[Contours of 3C~454.3.]{Dot-dashed curve: AGILE 95\% maximum 
      likelihood contour level; solid curve: AGILE 95\% error circle taking
      into account both statistic and systematic uncertainties; dotted
      curve: EGRET 95\% error circle \citep{Hartman1999:3eg}; 
      star: AGILE 95\% maximum 
      likelihood contour level barycentre; triangle: EGRET gamma-ray
      \source{} position \citep{Hartman1999:3eg}; 
      square: \source{} radio position.
    \label{3c454:fig:contour}}
\end{figure}

Figure~\ref{3c454:fig:map} shows a Gaussian-smoothed counts map 
($\sim 38^{\circ} \times 28^{\circ}$) in Galactic coordinates
integrated over the whole observing period, using the selections described 
in~\S\ref{3c454:dataanal}.
The source detection significance is 
13.8-$\sigma$ as derived from a maximum likelihood analysis.
The \gray source detected by the AGILE--GRID is fully
consistent with the radio position of \source{}, as shown in
Figure~\ref{3c454:fig:contour}. 
The solid curve represents the AGILE error circle taking 
into account both statistic and systematic uncertainties, 
the dot-dashed curve represents the 95\% maximum likelihood contour level,
while the dotted curve is the Third EGRET Catalog 95\% error circle 
(radius $0.\!^{\circ}28$).
Whilst the AGILE 95\% maximum likelihood contour level has a 
semi-major axis $a=0.\!^{\circ}33$ and a 
semi-minor axis $b=0.\!^{\circ}32$, the overall AGILE error circle
has a radius $r=0.\!^{\circ}47$.
The distance between the \source{} radio position (square)
and the AGILE 95\% maximum likelihood contour level barycentre
(star; $l=86.\!^{\circ}42$\,,\,$b=-38.\!^{\circ}37$) is
$\sim 0.\!^{\circ}29$. 
We note that the EGRET 95\% error circle corresponds to the
sum of all the pointings towards the source 
(VP1234, April 1991 - October 1995).
During this period EGRET accumulated 467 counts above 100~MeV,
while AGILE, during its 1-week pointing, accumulated 107 counts above
100~MeV. The much larger period over which the EGRET source
location was derived explains the relative sizes of the EGRET
and AGILE error circles.

\begin{figure}[!ht] 
    \includegraphics[angle=0,scale=0.45]{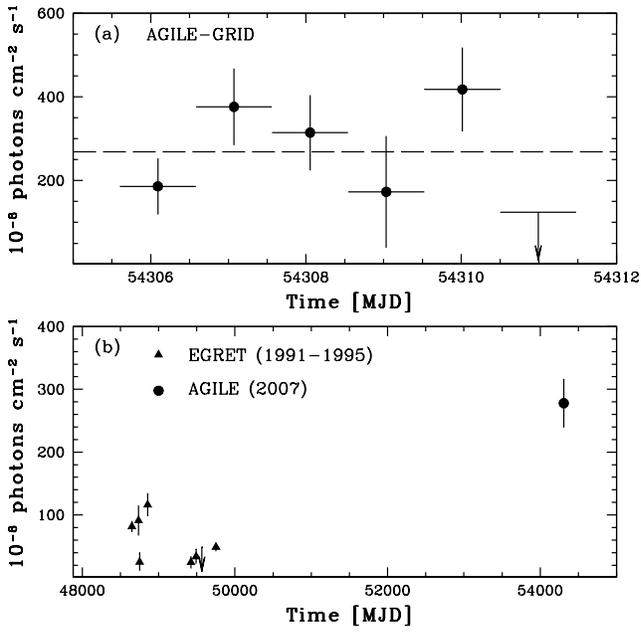}
    \caption[agile lc]{
      (a): AGILE--GRID \gray light-curve at 
       $\approx 1$-day resolution for E$>$100~MeV in units of 
      $10^{-8}$\,\phcmsec. The downward arrow represents a 2-$\sigma$
      upper-limit. The dashed line represents the weighted mean flux.
      (b): EGRET (triangles) and 
      AGILE--GRID (circle) gamma-ray
      lightcurve in units of $10^{-8}$\,\phcmsec. 
      EGRET data are from \citet{Hartman1999:3eg}.
    \label{3c454:fig:lcagile}}
\end{figure}

Figure~\ref{3c454:fig:lcagile}, panel (a), shows the \gray light-curve
at 1-day resolution for photons above 100~MeV. 
We notice that \source{} is detected at a 4-$\sigma$ level
during almost the whole period on a 1-day timescale; this clearly
indicates strong \gray flaring activity. The 
average \gray flux above 100 MeV for the whole period is
$ F_{\rm 3C~454.3}^{\rm est} = (280 \pm 40) \times 10^{-8}$\,\phcmsec.
The spectral analysis will be presented in a forthcoming paper.

The average \gray flux as well as the daily values
of the six days were derived according to \citet{Mattox1993:1633}.
First, the entire period was analyzed to determine the diffuse
gas parameters and then the source flux density was estimated
independently for each of six 1-day periods with the diffuse
parameters fixed at the values obtained in the first step.
The average \gray flux above 100 MeV for the whole period is 
the highest ever detected from this
source, as shown in Figure~\ref{3c454:fig:lcagile}, (b).
Fitting the GRID fluxes to a constant model (the weighted
mean of the 1-day average flux values) yields 
$\chi^2=2.611$ for 5 degrees of freedom (d.o.f.);
therefore we can exclude that the fluxes are constant at the 97.7\%
($\sim 2.3 \sigma$) level.
A more robust assessment of the source behaviour will be
possible when the calibrations are finalized. 
However, an indication can be seen of a quick flux decay
towards the end of the observation.

The source was not detected (above 5-$\sigma$) by the Super-AGILE
Iterative Removal Of Sources (IROS) applied to the $Z$ image, in
the 20--60~keV energy range. An upper limit to the observed count
rate was obtained by a study of the background fluctuations at the
position of the source and a simulation of the source and background
contributions with IROS. 
Assuming a power law spectral shape with photon index $\Gamma=1.5$ 
(extrapolated from the spectral fit of the simultaneous Swift 0.3-10 keV 
data, whose analysis will be presented elsewhere), we
find a 3-$\sigma$ upper limit of $2.3 \times 10^{-3}$\,\phcmsec{}
on the average flux from \source{}, allowing for a 50\% systematic error
to take into account the early status of the calibration of the 
Super-AGILE response.

         \section{Discussion} \label{3c454:discussion}

In this paper we presented the first Blazar detection by AGILE.
Previous EGRET observations and detections 
of \source{} were carried out in different observational 
conditions (off-axis angle between $7^{\circ}$ and 
$23^{\circ}$; typical viewing period duration 8 days, with
a single long observation lasting approximately 19 days, 
$7^{\circ}$ off-axis).
A significant \gray flare was detected in 1992 January-February 
with a peak level of $F_{\rm E>100MeV} = 140 \times 10^{-8}$\,
photons~cm$^{-2}$~s$^{-1}$ followed by a 2-week decreasing trend
\citep{Hartman1993:3C454_EGRET}. 
Subsequent EGRET detections in 1992 April-May confirmed the flux level near 
$F_{\rm E>100MeV} = 100 \times 10^{-8}$\,
photons~cm$^{-2}$~s$^{-1}$\, while an observation during 1995
November-December detected a significantly lower flux,
near $F_{\rm E>100MeV} = (20 - 30) \times 10^{-8}$\,
photons~cm$^{-2}$~s$^{-1}$ \citep{Aller1997:3C454_EGRET}.

The AGILE detection therefore shows the highest \gray
flux ever detected from \source.
It is remarkable that a FSRQ such as
\source{} is quite active in the high-energy domain over a
timescale of 2 decades. It is at the moment unclear what
determines the jet activity and the high-energy emission
from the environment close to the massive black hole of
\source.
In the EGRET era, nine blazars were observed to reach
a \gray flux (in units of $\times 10^{-8}$\,\phcmsec\, 
above 100~MeV) between 100--200, two (3C~279 and 1622$-$297) 
between 200--300, and two (0528$+$134 and 1622$-$297) between 
300--350. 
Thus, our detection of \source{} is comparable to the highest 
\gray fluxes detected so far.

The AGILE \gray detection of \source{} supports
the idea that a special class of
blazars shows significant \gray activity on 
timescales of decades that clearly distinguishes them from other
blazar candidates that share similar
radio, optical, or X-ray properties,
but show no \gray emission.
What determines such behaviour, observed in \source{}, 3C~279 and
similar \gray blazars, is currently unclear.
\citet{Kellermann2004:jetvel} find that radio jets of strong
\gray blazars seem to have significantly faster median apparent speeds,
$(8.6 \pm 1.6)c$\,, than blazars with no \gray emission,
$(3.9 \pm 1.1)c$. This implies that \gray sources have larger
Doppler factors than blazars with no \gray emission.
Recently, \citet{Kovalev2005:compact} found that \gray
blazar radio jets appear to have more compact VLBA structure
than blazars with no \gray emission, 
contrary to earlier results \citep{Kellermann1998:firstradio} wherein
the authors reported no clear differences in the jet
morphology between \gray loud and \gray quiet blazars.
Because of its large field
of view and relatively flat sensitivity in the \gray
energy range, AGILE can, for the first time,
simultaneously monitor a large number of known and
candidate \gray blazars, both in \gray and
in the hard X-ray energy bands.
This will make it possible to study in detail the apparent dichotomy 
in the blazar population between \gray loud and \gray 
quiet sources.
Moreover, blazar \gray light-curves can now be monitored on 
timescales of the order of 3 weeks, allowing us to perform long-term 
variability studies both in the high-energy range, and correlated 
with the optical band. This will be crucial in investigating
correlations and time delays between optical flares and \gray
emission. 

\acknowledgments
We thank the Referee for his/her very constructive comments.
The AGILE Mission is funded by the Italian Space Agency (ASI) with
scientific and programmatic participation by the Italian Institute
of Astrophysics (INAF) and the Italian Institute of Nuclear
Physics (INFN). This investigation was carried out with partial
support under ASI contract N. {I/089/06/1}.
We warmly thank G. Tosti for his restless monitoring efforts.
We wish to express our gratitude to
the Carlo Gavazzi Space, Thales Alenia Space, Telespazio and ASDC/Dataspazio 
Teams that implemented the necessary procedures to carry out the
AGILE re-pointing.

{\it Facilities:} \facility{AGILE}.


\clearpage

%
%
\begin{deluxetable}{lll} 
  \tablecolumns{3}
  \tabletypesize{\scriptsize}
  \tablecaption{Gamma--Ray Imaging Detector and Super--AGILE 
    Characteristics\label{3c454:tab:tab1}}
  \tablewidth{0pt}
  \tablehead{
    \colhead{Parameter} & \colhead{AGILE--GRID}  & \colhead{Super--AGILE}}
  \startdata
  Energy Range          & 30~MeV--50~GeV & 18--60~keV \\
  Field of view         & $\sim 3$\,sr  
                          & $2\times(107^{\circ} \times 68^{\circ})$\tablenotemark{a}\\
  Sensitivity           & $3 \times 10^{-7}$\phcmsec \tablenotemark{b} 
                          & $\sim 15$\,mCrab\tablenotemark{c} \\
  Angular Resolution    & $1.2^{\circ}$ \tablenotemark{d}   
                          & $6$\,arcmin\tablenotemark{e}  \\
  Source Loc. Accuracy  & $\sim 15$\,arcmin\tablenotemark{f} 
                          & $\sim$ 2--3\,arcmin \tablenotemark{g}\\
  Energy resol.         & $\Delta E/E \sim 1$\tablenotemark{h} 
                          & $\Delta E = 8$\,keV \tablenotemark{i}\\
  Absolute time resol.  & $\sim 2$\,$\mu$s & $\sim 5$\,$\mu$s\\
  \enddata
  \tablenotetext{(a)}{ FOV of each half detector (Full Width at Zero Intensity).}
  \tablenotetext{(b)}{ E $>$\,100~MeV, 
    $5\sigma$ in $10^{6}$\,s at high galactic latitude, on axis.}
  \tablenotetext{(c)}{ $5\sigma$ in $50$\,ks, on axis.}
  \tablenotetext{(d)}{ 68\% containment radius at 400~MeV.}
  \tablenotetext{(e)}{ Sky pixel size on-axis.}
  \tablenotetext{(f)}{ S/N $\sim 10$\,; 90\% confidence limit (c.l.) radius
    at high galactic latitude. Statistical only.}
  \tablenotetext{(g)}{ For sources at $10\sigma$\,, including statistics and systematics.}
  \tablenotetext{(h)}{ At 400~MeV.}
  \tablenotetext{(i)}{ FWHM.}
\end{deluxetable}

\clearpage

%
%
%

%
%
%

%
%
%

%

\end{document}